\documentclass[aps,prx,twocolumn,notitlepage,showkeys,showpacs,superscriptaddress,longbibliography,floatfix]{revtex4-1}
\usepackage{epsf}
\usepackage{framed}
\usepackage{caption}
\usepackage{bm}
\usepackage{times}
\usepackage{amsfonts}
\usepackage{amssymb}
\usepackage{amsmath,mathtools}
\usepackage{array}
\usepackage{enumerate,dsfont,here,url}
\usepackage{dcolumn,multirow}
\usepackage{slashbox}
\usepackage{bbding}
\usepackage{tabularx}
\usepackage[colorlinks=true,citecolor=blue,linkcolor=blue]{hyperref}
\usepackage[normalem]{ulem}
\usepackage{makecell}

\begin{document}
\setcounter{figure}{0}
 \captionsetup[figure]{name={\textbf{Fig.}},labelsep=period,singlelinecheck=off,justification=raggedright}
 \setcounter{table}{0}
 \captionsetup[table]{name={\textbf{Table}},labelsep=period,singlelinecheck=off,justification=raggedright}
\title{Topological state evolution by symmetry-breaking}
\author{Feng Tang}\email{fengtang@nju.edu.cn}
\author{Xiangang Wan}
\affiliation{National Laboratory of Solid State Microstructures and School of Physics, Nanjing University, Nanjing 210093, China and Collaborative Innovation Center of Advanced Microstructures, Nanjing University, Nanjing 210093, China}
\begin{abstract}
Previous symmetry-based database searches have already revealed ubiquitous band topology in nature, while the destiny of band topology under symmetry-breaking is yet to be studied comprehensively.  Here we first develop a framework  allowing systematically ascertaining topological state evolution as expressed via a tree-like graph for magnetic/non-magnetic crystalline material belonging to any of the 1651 magnetic space groups. Interestingly, we find that specifying different ways of realizing symmetry-breaking leads to various contractions of the tree-like graph, as a new angle of comprehensively characterizing the correlation between a spontaneous symmetry-breaking and any symmetry-group-indicated physics consequence. We also  perform a  high-throughput investigation  on the 1267 stoichiometric magnetic materials ever-experimentally synthesized  to reveal a  hierarchy of topological states  along all continuous paths of symmetry-breaking (preserving the translation symmetry) from the parent magnetic space group to P1.  The results in this work are expected to aid  experimentalists  in selecting feasible and appropriate means to tune band topology towards realistic applications and promote further studies on using tree-like graph to explore the interconnection between topology and other intriguing orderings.
\end{abstract}
\date{\today}
\maketitle
\section{Introduction}
The past nearly two decades have witnessed  tremendous advancements of symmetry-protected topological  phases and topological materials \cite{MT-SM-RMP-Kane,MT-SM-RMP-Qi,MT-RMP-Chiu,MT-SM-TCI-Ando,Balatsky,MT-SM-RMP-AV,MT-Binghai-Review,MT-Wieder-Review,MT-tokura-review,MT-SM-Nature-Review-BAB} reshaping our fundamental understanding on the electronic properties in solids. In the well-established paradigm of studying topological phases in condensed matter, one usually utilizes the symmetry to classify the protected topological phases, as the ten-fold way does \cite{tenfold-1,tenfold-2,tenfold-3}.  Very recently,   topological quantum chemistry (TQC) \cite{MT-SM-TQC,MT-SM-M-TQC} or symmetry-indicators (SIs) \cite{MT-SM-SI,MT-SM-M-SI},  have been applied  in  discovering topological nonmagnetic \cite{MT-SM-N-1,MT-SM-N-2,MT-SM-N-3,MT-SM-AllTopo}/magnetic \cite{MT-SM-N-4}/superconducting \cite{MT-SM-Tang-TNSC} materials routinely using the first-principles calculated   high-symmetry point (HSP) symmetry-data with respect to the space groups/magnetic space groups (MSGs) of the pristine materials. These database searches  reveal  the ubiquitous band topology in nature, protected by the  MSG symmetry that describes the symmetry of spatial arrangement of atoms/ions (and magnetic structure).

On the other hand, symmetry-breaking can be utilized in creating new topological phase. For example, quantum anomalous Hall effect was realized  through introducing mass term in the time-reversal symmetry protected surface Dirac fermion by  magnetic dopant \cite{MT-RuiYu,MT-SM-QAH-Bi2Te3-film}.  Unconventional fermions can emerge in crystals which break the  Poincar$\acute{\mathrm{e}}$ symmetry \cite{MT-SM-NewFermions}. And a higher-order band topology is induced by applying strain on SnTe to break the mirror symmetry which originally protects a mirror-symmetry protected topological crystalline insulating phase \cite{MT-SM-H-Topo-2}. Furthermore, symmetry-breaking due to external fields could  act as a versatile controlling knob on  topological states since such symmetry-breaking can be easily-manipulated, while that spontaneously caused by a phase transition could bridge the corresponding ordering and related band topology.

Even though concrete  symmetry-breaking  cannot be exhausted, we can classify  symmetry-breaking by group-subgroup relations so that  various meticulous concrete  symmetry-breaking ways corresponding to one subgroup are anticipated to share a common topological diagnosing prediction,  since only the transformation properties of HSP wave-functions account \cite{MT-SM-SI,MT-SM-Tang-NP,MT-SM-M-SI}. Different from  the group-subgroup relations for the 32 point groups well-tabulated in many group-theory textbooks, those for the 1651 MSGs have not yet been obtained.   In this work, we first construct all subgroups preserving the original translation symmetry (called t-subgroups)  based on the 1651 MSGs, compute the compatibility relations between the HSPs of the parent MSG and the corresponding $k$ points for each t-subgroup, laying a foundation of calculating topological state evolution (TSE) by symmetry-breaking using SIs \cite{MT-SM-SI,MT-SM-Tang-NP,MT-SM-M-SI}. We express TSE by a tree-like graph showing t-subgroups as vertexes  and symmetry-breaking paths by edges (see Fig. \ref{figure-1}).  We also perform high-throughput calculations on the realistic magnetic materials as listed in MAGNDATA \cite{MT-SM-magndata}, which have been synthesized and whose magnetic structures have  already been  deduced  experimentally.

Owing to the advanced techniques on ultrafast light and high magnetic fields, experimentalists  are able to apply controllable external stimuli to pristine crystals via symmetry-breaking  to tune the topological states therein guided by the results of this work. Interestingly, the topological state can be drastically modified, thus leading to giant quantum responses which have been observed in realistic materials, such as the colossal angular magnetoresistance in ferrimagnetic Mn$_3$Si$_2$Te$_6$ \cite{MT-SM-Mn2Si3Te6-N, MT-SM-Mn2Si3Te6-N-2,MT-SM-Mn2Si3Te6-PRB} and  the giant anomalous Hall conductivity in Heusler Co$_2$MnAl \cite{MT-SM-CoMnAl}, where the topological states are controlled by  rotating the magnetic moments.
Hence we  expect our results to aid in unveiling more magnetic topological materials with  unprecedentedly fascinating properties driven by the change of topological phase.

Interestingly, we also find that by specifying concrete realization of symmetry-breaking, the tree-like graph would be encountered with a contraction (see the dashed circles in Fig. \ref{figure-1}(a)) and transform to a different tree-like graph and TSE concomitantly. Since the symmetry-breaking could be induced by some ordering  spontaneously, this finding  then lays a new framework of studying the correlation among various types of orderings and symmetry-group-indicated physical quantity based on the contracted tree-like graph.

\section{Results}
\begin{figure*}[!hbtp]
  \includegraphics[width=1\textwidth]{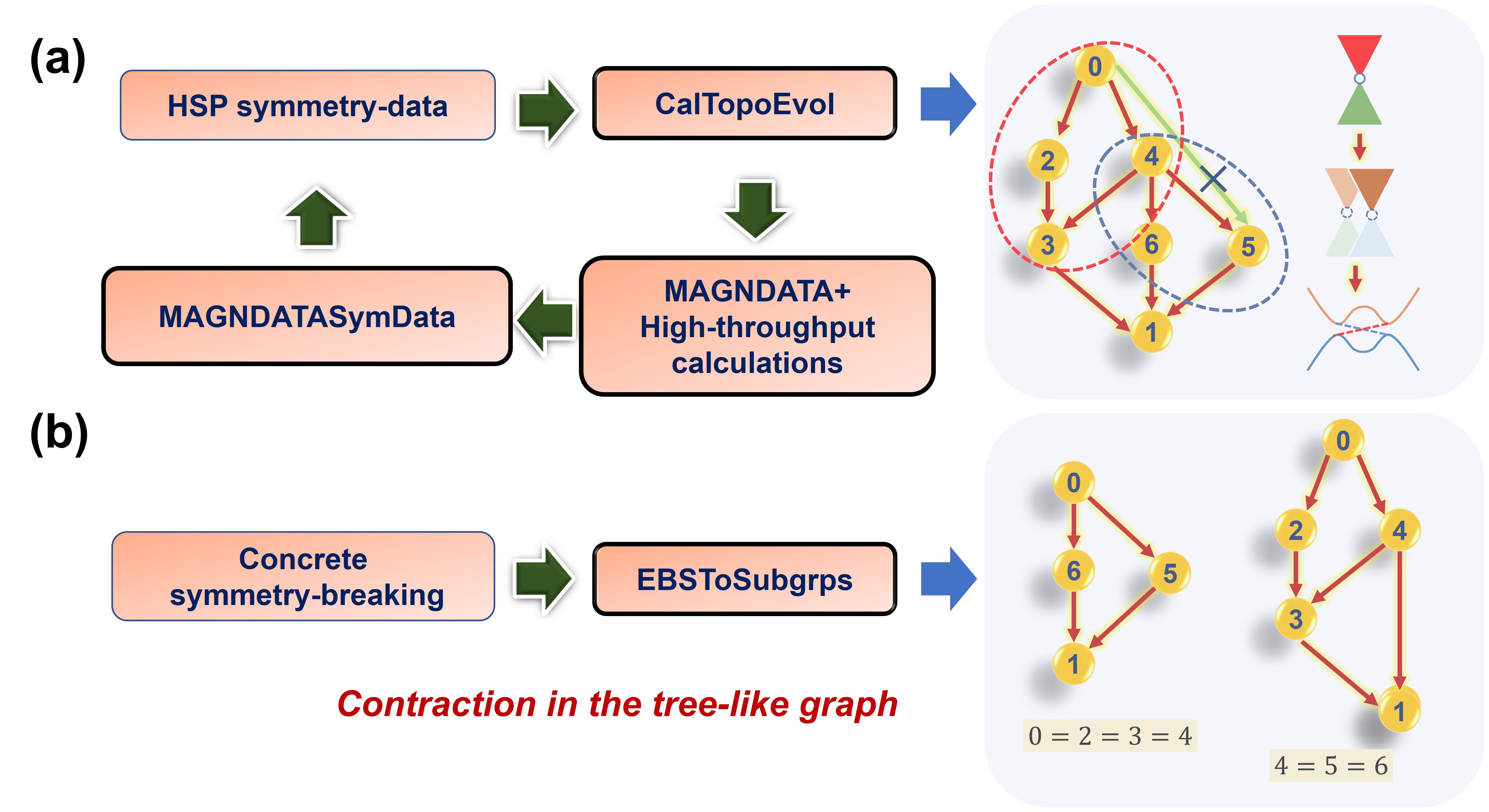}\\
  \caption{\textbf{Main results.} (a) Given HSP symmetry-data, namely, the set of counts of occurrences of (co-)irreps at HSPs, CalTopoEvol outputs a tree-like graph including all t-subgroups represented by vertexes of the graph. Clicking on each vertex, a window popups showing detailed topological state diagnosed by SI method. In the right panel, we show an example that topological state evolves from Dirac semimetal to Weyl semimetal, and then becomes a topological insulator. (b) Note that we also integrate our high-throughput calculation results in a subroutine called ``MAGNDATASymData'', which contains all successfully computed magnetic materials in MAGNDATA \cite{MT-SM-magndata} considering different values of $U$ and  varying the electron filling ($\nu$) by $|\nu-\nu_0|\le 4$, where $\nu_0$ denotes the intrinsic filling. (b) We also develop a subroutine called ``EBSToSubgrps'', which consider seven external fields for each MSG to realize t-subgroups. Note that given a concrete symmetry-breaking, the  tree-like graph might only be contracted from the one shown in (a): ``$0=2=3=4$'' and ``4=5=6'' denote two contractions corresponding to red and blue dashed ellipses in (a). }\label{figure-1}
\end{figure*}

We first summarize our results in this part. The results are categorized into three parts: A framework of computing TSE using SIs by tree-like graph, high-throughput calculations on magnetic materials and the contraction by concrete symmetry-breaking in tree-like graph. As schematically shown in Fig. \ref{figure-1}, corresponding to these parts, we develop three subroutines, respectively, called: ``CalTopoEvol'', ``MAGNDATASymData'' and ``EBSToSubgrps''. Below, we present a brief description and leave a more detailed description to Supplementary Material (SM) \cite{MT-seeSM}.

\subsection{Computing TSE using SIs by tree-like graph: CalTopoEvol}
In order to find TSE by symmetry-breaking, one need to exhaust all subgroups and then depict a tree-like graph (see Fig. \ref{figure-1} for examples) to demonstrate the symmetry-breaking paths: Each vertex of the tree-like graph denotes one subgroup, and it is required that one cannot find another subgroup between any pair of vertexes along any path. In this way, all subgroups are included with no omission in any symmetry-breaking path. For example, as shown in Fig. \ref{figure-1}(a), there are 7 subgroups, denoted by $0,1,2,\ldots,6$ and there are 4 symmetry-breaking paths: $0\rightarrow 2\rightarrow 3\rightarrow 1$, $0\rightarrow 4\rightarrow 3\rightarrow 1$, $0\rightarrow 4\rightarrow 6\rightarrow 1$ and $0\rightarrow 4\rightarrow 5\rightarrow 1$ (each arrow connects two subgroups from a higher symmetry group to a lower one). However, $0\rightarrow 5\rightarrow 1$ is not a symmetry-breaking path since $4$ is between 0 and 5.   Here we focus on the t-subgroups for MSGs. Enumerating all t-subgroups is equivalent to finding all subgroups of magnetic point group for a given MSG. We denote each t-subgroup ($G'$) by an integer: 0,1,2,\ldots (called symmetry-breaking pattern), and we always let 0 and 1 be the parent MSG ($G$) and P1 respectively.  In total, 63 762 t-subgroups are found for the 1651 MSGs. Note that any t-subgroup is still an MSG, meaning that the atomic-insulator basis sets \cite{MT-SM-SI,MT-SM-M-SI,MT-SM-Tang-NP} and $k\cdot p$ models around band crossings (BCs) \cite{SM-Tang-KP,MT-SM-Tang-BN} for the 1651 MSGs still apply. Besides, note that the HSP of $G'$ is definitely an HSP of $G$, meaning that the HSP symmetry-data for the HSPs of $G$ are sufficient for the diagnosis of band topology with respect to $G'$. Denote the HSP of $G$ as $\vec{k}$ and the corresponding $k$ point of $G'$ as $\vec{k}'$.  One can obtain the compatibility relations (CRs) between the irreducible representations (irreps) or irreducible co-representations (co-irreps) of little group of $\vec{k}$ and those of $\vec{k}'$:
 \begin{equation}\label{CR}
    D_{k}^j\rightarrow\sum_{j'}c_{jj'}D_{k'}^{j'},
 \end{equation}
 and thus,
   \begin{equation}\label{CR-n}
   n_{k'}^{j'}=\sum_{j} c_{jj'}n_{k}^j,
   \end{equation}
  where  $n_{k/k}^{j/j'}$ denotes the number of occurrences of (co-)irrep $D_k^j/D_{k'}^{j'}$ at $\vec{k}/\vec{k}'$.  Eq. \ref{CR} can be computed before realistic materials calculations. It is worth pointing out that once an HSP is partially occupied (namely, not all $(n_k^j)^,$s  are  integers), Eq. \ref{CR-n} might not hold and one need to consider a permutation of $\{j'\}^,$s that are originated from $j$ (assuming $n_k^j$ is not an integer), as clarified in Sec. S2 of SM \cite{MT-seeSM}.

We then use symmetry-data for $G'$, namely $\{n_{k'}^{j'}\}$, to scan all high-symmetry lines/planes (HSLs/HSPLs) of $G'$ to detect enforced BCs. Note that $\vec{k}'$ might not be an HSP for $G'$. Consider an HSL/HSPL of $G'$, $\vec{\mathcal{K}'}$. Denote two adjacent $k$ points of $G'$ lying in the HSL/HSPL as $\vec{k}'_1$ and $\vec{k}'_2$ that are originated from two HSPs of $G$, and the symmetry-data are denoted by $n_{k_1'}^{j_1'}$ and $n_{k_2'}^{j_2'}$, respectively.  The CRs from $\vec{k}_1/\vec{k}_2$ to $\vec{\mathcal{K}'}$ are:
 \begin{equation}\label{CR2}
 \begin{aligned}
    &D_{k'_1}^{j_1'}\rightarrow\sum_{j''}d_{j'_1 j''}D_{\mathcal{K}'}^{j''},\\
    &D_{k'_2}^{j_2'}\rightarrow\sum_{j''}d_{j'_2 j''}D_{\mathcal{K}'}^{j''},
 \end{aligned}
 \end{equation}
and then one can obtain the symmetry-data of $\vec{\mathcal{K}'}$ from both sides:  $n_{\mathcal{K}'}^{j''}=\sum_{j_1'} d_{j'_1j''}n_{k_1'}^{j_1'}$ and  $n_{\mathcal{K}'}^{j''}=\sum_{j_2'} d_{j'_2j''}n_{k_2'}^{j_1'}$. Once they are not equal, there exists enforced BC somewhere between $\vec{k}'_1$ and $\vec{k}'_2$ along $\vec{\mathcal{K}'}$.  Once no enforced BCs are found after scanning all HSLs/HSPLs of $G'$, the symmetry-data at HSPs of $G'$ are then utilized to be expanded on the atomic insulator basis for $G'$ to diagnose whether it owns a nontrivial SI. Note that $n_{k_1'}^{j_1'}$ and $n_{k_2'}^{j_2'}$ need to be all integers otherwise BC should be pinned at the corresponding $k$ point (say $n_{k_1'}^{j_1'}$ is not an integer, BC then appears at $\vec{k_1}'$ carrying $D_{k_1'}^{j_1'}$).  All the above described steps have already be integrated into CalTopoEvol, and one only need input the parent MSG (in the Belov-Neronova-Smirnova notation \cite{MT-SM-BNS}) and the corresponding HSP symmetry-data (namely, $\{n_k^j\}$). See examples of using CalTopoEvol in Sec. S2 of SM \cite{MT-seeSM}. To make CalTopoEvol used more conveniently, a subroutine called ``ToMSGKpoint'' was recently developed, capable of outputting the inputs of CalTopoEvol given any customized nonmagnetic/magnetic crystal structure \cite{ToMSGKpoint}.

\subsection{High-throughput calculations: MAGNDATASymData}
 Of the 1721 magnetic materials in total which we collected at the time of initiating this work from MAGNDATA \cite{MT-SM-magndata}, we first filter out materials with incommensurate magnetic ordering,  requiring  that the atoms fully occupy  their sites, the lattice parameters  are compatible with the experimentally identified MSG and the numbers of ions  are compatible with the composition dictated by the chemical formula. We  finally obtain 1267 ``high-quality'' magnetic materials as our starting point for subsequent first-principles calculations. The experimentally identified MSGs \cite{MT-SM-magndata} are set as the parent MSG.  The electron correlation is considered by choosing various values of  Hubbard $U$, and thus we have 5883 jobs in total for the 1267 materials (one material with one value of $U$ is counted as one job) (see Sec. S6 of SM \cite{MT-seeSM} for details). Finally, 5062 jobs reach convergence in the first-principles calculations, and output the convergent magnetic moments. In Sec. S9 of SM \cite{MT-seeSM}, we tabulate all the calculation results for all the 1267 magnetic materials, where comparison of first-principles calculated and experimental magnetic moments and plots of energy bands and density of states (DOS) can be found. Further ensuring that the converged magnetic structure  preserves the original MSG symmetry, we then have 4012 jobs and for these jobs, we compute the representations for all the energy levels at the HSPs for the parent MSG  by the first-principles calculated wave-functions that are then utilized to compute HSP symmetry-data (see Sec. S6 of SM for details \cite{MT-seeSM}). These jobs  correspond to 295 MSGs and 1013 magnetic materials in total.  In order to detect all topological bands around the Fermi level, we vary the electron filling to be $\nu=\nu_0, \nu_0\pm1,\nu_0\pm2,\nu_0\pm3,\nu_0\pm4$ ($\nu_0$ denotes the intrinsic filling, which is the number of valence electrons per primitive unit cell). Note that tuning filling might also lead to different topological states, as has been realized in a topological phase change transistor by electrostatic gating \cite{MT-transistor}.  All these HSP symmetry-data can be retrieved by the subroutine, MAGNDATASymData and can be used as the input of CalTopoEvol to get detailed TSE.  See Sec. S3 of SM \cite{MT-seeSM} for examples of using MAGNDATASymData.

\subsection{Contraction in tree-like graph: EBSToSubgrps}
It is worth pointing out that there might exist versatile concrete ways of symmetry-breaking leading to the same t-subgroup, thus sharing the same topological state. Enumerating all t-subgroups and then depicting the associated tree-like graph could exhaust all TSEs along all symmetry-breaking paths, irrespective of how symmetry-breaking is imposed concretely, owing to the fact that the band topology can be diagnosed by HSP symmetry-data \cite{MT-SM-SI,MT-SM-M-SI,MT-SM-Tang-NP}. However,  once restricting concrete way of symmetry-breaking, it is interesting to find that the corresponding tree-like graph could be contracted to demonstrate a different pattern, as shown in Fig. \ref{figure-1}(b), where two different contractions lead to two  different contracted tree-like graphs.  Here we consider seven different ways of symmetry-breaking and all of them are uniformly applied to the pristine material  that would not break the translation symmetry, including:
\begin{equation}\notag
\begin{aligned}
&\text{1. Only applying electric field,}\\
&\text{2. Only applying magnetic field,}\\
&\text{3. Only applying strain field,}\\
&\text{4. Simultaneously applying electric and magnetic fields,}\\
&\text{5. Simultaneously applying electric and strain fields,}\\
&\text{6. Simultaneously applying magnetic and strain fields,}\\
&\text{7. Simultaneously applying electric, magnetic and strain fields.}
\end{aligned}
\end{equation}

Accordingly, we have developed a subroutine, EBSToSubgrps, which shows the resulting contracted tree-like graph (where the applied fields and the preserved operations are indicated in the vertexes) given an MSG and specifying concrete way of symmetry-breaking.  See examples of using EBSToSubgrps in Sec. S4 of SM \cite{MT-seeSM}. Note that the above fields can also be understood as orderings falling into the same symmetry class. For example, the electric and magnetic fields correspond to macroscopic electric polarization and magnetization, respectively.   We will show concrete examples in Sec. \ref{Mat-example} by MSGs 12.62 and 126.386.

\section{Selected magnetic topological materials}
Since the parameter Hubbard $U$ could change the topology prediction, we firstly select 236 calculated magnetic materials that are predicted to be topological for at least 3 different values of Hubbard $U$, whose ID$^,$s in MAGNDATA  \cite{MT-SM-magndata} are listed in Table \ref{table-1}. It is also found that the number of magnetic topological materials hosting such robustness against $U$ is gradually increased to 725 when the fillings satisfying $|\Delta\nu|\le4$ ($\Delta\nu=\nu-\nu_0$) as shown in Fig. S10 of SM \cite{MT-seeSM}, indicating that more than 70\% of the investigated magnetic materials can be topological.

\begin{table*}[!t]
 \resizebox{1.\textwidth}{!}{\begin{tabular}{l|l|l|l|l|l|l|l|l|l}
 		\hline\hline
 ID:formula&ID:formula&ID:formula&ID:formula&ID:formula&ID:formula&ID:formula&ID:formula&ID:formula&ID:formula\\
 		\hline
 {\color{red}0.108:$\text{Mn}_3\text{Ir}$}&{\color{red}0.109:$\text{Mn}_3\text{Pt}$}&0.126:$\text{NpCo}_2$&0.140:$\text{LuFe}_4\text{Ge}_2 $&0.149:$\text{Nd}_3\text{Ru}_4\text{Al}_{12}$&0.174:$\text{Pr}_3\text{Ru}_4\text{Al}_{12}$&{\color{red}0.177:$\text{Mn}_3\text{GaN}$}&0.186:CeMnAsO&0.194:$	 \text{UPt}_2\text{Si}_2 $&0.199:$\text{Mn}_3\text{Sn}$\\
 		\hline
      	0.200:$\text{Mn}_3\text{Sn}$&0.203:$\text{Mn}_3\text{Ge}$&0.207:$  \text{TlFe}_{1.6}\text{Se}_{2}$&0.227:$\text{NdCo}_2$&{\color{red}0.228:$ \text{TbCo}_2$}&0.236:$ \text{CaFe}_4\text{Al}_8$&0.27:$\text{YFe}_4\text{Ge}_2$&{\color{red}0.273:$\text{Mn}_3\text{ZnN}$}&0.274:$\text{Mn}_4\text{N}$&{\color{red}0.275:$\text{Mn}_3\text{AlN}$}\\
        \hline
         	 0.276:$\text{Mn}_3\text{AlN}$&0.279::$\text{Mn}_3\text{As}$&0.280:$\text{Mn}_3\text{As}$&0.286:$\text{Mn}_5\text{Ge}_3$&0.320:$\text{U}_2\text{Pd}_2\text{In}$&0.321:$\text{U}_2\text{Pd}_2\text{Sn}$&0.325:$\text{Cd}\text{Yb}_2\text{Se}_4$&0.327:$\text{CsMnF}_4$&0.365:$\text{Ba}\text{Cr}_2\text{As}_4$ &0.367:$\text{Eu}\text{Cr}_2\text{As}_4$\\
         \hline
         	0.374:$ \text{YNi}_4\text{Si}$&0.377:Mn$_3$Ge &0.378:UBi$_2$&0.395:MnPtGa&0.396:MnPtGa&0.402:Sr$_4$Fe$_4$O$_{11}$&0.407:NdSi&0.408:PrSi&0.414:AlFe$_2$B$_2$&0.415:EuFe$_2$P$_2$\\
         \hline
         	 0.436:TbNi$_4$Si&0.445:MnCoGe&0.454:PrScSb&0.461:CoRh$_2$O4&0.473:LaMn$_2$Si$_2$&0.486:ErCr$_2$Si$_2$&0.487:ErCr$_2$Si$_2$&0.495:LaMn$_2$Si$_2$&0.496:LaMn$_2$Si$_2$&0.497:LaMn$_2$Si$_2$\\
         \hline
         	 0.512:Mn$_3$As$_2$&0.561:NdNiGe$_2$&0.566:TbNiGe$_2$&0.593:UPSe&0.594:UAsS&0.595:UPTe&0.596:UAsTe&0.599:CaMnSi&0.600:CaMnSi&0.603:CaMn$_2$Ge$_2$\\
         \hline
         	 0.604:CaMn$_2$Ge$_2$&0.605:BaMn$_2$Ge$_2$&0.606:BaMn$_2$Ge$_2$&0.609:NdMnO$_3$&0.613:FeCr$_2$S$_4$&0.614:FeCr$_2$S$_4$&0.615:FeCr$_2$S$_4$&0.616:HoB$_2$&0.623:NdMnAsO&0.625:U$_2$Pd$_2$In\\
         \hline
        0.641:Mn$_3$Ga & 0.656:NdMn$_2$Ge$_2$ & 0.657:PrMn$_2$Ge$_2$ & 0.662:Mn$_3$Sn$_2$ & 0.663:Mn$_3$Sn$_2$ & 0.664:Mn$_3$Sn$_2$ & 0.665:CeMnSbO & 0.681:Ce4Sb$_3$ & 0.684:TbPt & {\color{red}0.699:LiMn$_6$Sn$_6$} \\
        \hline
        {\color{red}0.702:TbMn$_6$Sn$_6$} & {\color{red}0.703:HoMn$_6$Sn$_6$} & 0.706:Tb$_2$Ir$_3$Ga$_9$ & 0.707:Tb$_2$Ir$_3$Ga$_9$ & 0.709:MnNb$_4$S$_8$ & 0.711:MnTa$_4$S$_8$ & 0.732:SrRuO$_3$ & 0.737: LaBaMn$_2$O$_6$ & 0.738:LaBaMn$_2$O$_6$ & {\color{red}0.74:Mn$_3$Cu$_{0.5}$Ge$_{0.5}$N}  \\
        \hline
        0.747:Ba$_3$CoIr$_2$O$_9$ & 0.771:PrMnSi$_2$ & 0.772:PrMnSi$_2$ & 0.773:NdMnSi$_2$ & 0.774:NdMnSi$_2$ & 0.776:CeMnSi$_2$ & 0.777:CeMnSi$_2$ & 0.778:LaMnSi$_2$ & 0.779:LaMnSi$_2$ & 0.780:LaMnSi$_2$ \\
        \hline
        0.80:U$_2$Pd$_2$In  & 0.81:U$_2$Pd$_2$Sn  & 1.0.11:CeCoGe$_3$ & 1.0.12:UAu$_2$Si$_2$ & 1.0.29:CeIrGe$_3$ & 1.0.43:UPd$_2$Si$_2$ & 1.102:U$_2$Ni$_2$In & 1.103:U$_2$Rh$_2$Sn& 1.104:Gd$_2$CuO$_4$ & {\color{red}1.110:ScMn$_6$Ge$_6$}\\
        \hline
        1.130:Cr$_2$As& 1.131:Fe$_2$As & 1.132:Mn$_2$As & 1.139:Ho$_2$RhIn$_8$& 1.143:Mn$_3$Pt& 1.146:LaCrAsO & 1.150:PrAg & 1.16:BaFe$_2$As$_2$  & 1.160:UP & 1.162:NdMg\\
        \hline
        1.179:NdCoAsO & 1.187:TbRh$_2$Si$_2$& 1.188:CeRh$_2$Si$_2$& 1.207:U$_2$Rh$_2$Sn & 1.208:UAs & 1.21:DyCo$_2$Si$_2$  & 1.215:UP$_2$ & 1.222:Er$_2$CoGa$_8$ & 1.223:Tm$_2$CoGa$_8$ & 1.251:NdCo$_2$P$_2$\\
        \hline
        1.252:CaCo$_2$P$_2$ & 1.253:CeCo$_2$P$_2$ & 1.254:UNiGa$_5$ & 1.255:UPtGa$_5$& 1.261:NpRhGa$_5$ & 1.290:CeRh$_2$Si$_2$& 1.291:CeAu$_2$Si$_2$ & 1.296:PrNi$_2$B$_2$C & 1.305:Mn$_5$Si$_3$ & 1.308:MnBi$_2$Te$_4$ \\
        \hline
        1.338:U$_2$Ni$_2$In & 1.369:HFe$_2$Ge$_2$ & 1.384:USb$_2$ & 1.398:Pr$_2$CuO$_4$ & 1.399:Pr$_2$CuO$_4$ & 1.400:TbAg$_2$ & 1.419:GdIn$_3$ & 1.421:NdRh$_2$Si$_2$ & 1.422:ErRh$_2$Si$_2$ & 1.423:UPb$_3$\\
        \hline
        1.425:UGeTe & 1.427:HoCo$_2$Ge$_2$& 1.428:UN & 1.442:URu$_2$Si$_2$& 1.446:CeCoAl$_4$& 1.453:EuMn$_2$Si$_2$ & 1.458:CsCo$_2$Se$_2$ & 1.460:PrCuSi & 1.461:Sr$_2$Cr$_3$As$_2$O$_2$ & 1.468:TbMn$_2$Si$_2$\\
        \hline
        1.469:YMn$_2$Si$_2$ & 1.475:DyNiAl$_4$ & 1.479:U$_2$Ni$_2$Sn& 1.487:CeIrAl$_4$Si$_2$ & 1.488:CeMn$_2$Si$_2$ & 1.489:CeMn$_2$Si$_2$ & 1.490:CeMn$_2$Si$_2$ & 1.491:PrMn$_2$Si$_2$ & 1.492:PrMn$_2$Si$_2$& 1.493:NdMn$_2$Si$_2$\\
        \hline
        1.494: NdMn$_2$Si$_2$ & 1.495:YMn$_2$Si$_2$ & 1.496:YMn$_2$Ge$_2$ & 1.497:EuMg$_2$Bi$_2$ & 1.508:Mn$_2$AlB$_2$ & 1.512:TbCo$_2$Si$_2$ & 1.513:HoCo$_2$Si$_2$ & 1.514:HoCo$_2$Si$_2$& 1.516:ErCo$_2$Si$_2$ & 1.52:CaFe$_2$As$_2$\\
        \hline
        1.530:CeC$_2$ & 1.531:PrC$_2$& 1.532:NdC$_2$ & 1.534:HoC$_2$& 1.536:UPd$_2$Si$_2$ & 1.537:URh$_2$Si$_2$ & 1.549:U$_2$Ni$_2$In& 1.555:Mn$_3$B$_4$ & 1.558:MnSn$_2$ & 1.559:MnSn$_2$\\
        \hline
        1.568:GdCu$_2$Si$_2$ & 1.575:ErRh & 1.585:PrFeAsO & 1.623:EuMg$_2$Bi$_2$& 1.628:PrMnSi$_2$ & 1.635:ErFe$_2$Si$_2$ & 1.636:ErMn$_2$Si$_2$ & 1.637:ErMn$_2$Si$_2$ & 1.638:ErMn$_2$Ge$_2$ & 1.639:ErMn$_2$Ge$_2$ \\
        \hline
        1.640:ErMn$_2$Ge$_2$ & 1.80:Dy$_2$CoGa$_8$  & 1.81:GdIn$_3$  & 1.82:Nd$_2$RhLn$_8$ & 1.85:alpha-Mn & 1.87:Tb$_2$CoGa$_8$ & 1.88:Mn$_5$Si$_3$  &2.1:EuFe$_2$As$_2$ & 2.10:HoP  & 2.11:TbMg\\
        \hline
         2.12:TbMg  &2.13:UP & 2.14:NdMg & 2.19:Mn$_3$ZnC  & 2.26:PrCo$_2$P$_2$  & 2.28:NpNiGa$_5$  & 2.31:Mn$_3$ZnN & 2.48:Pr$_2$CuO$_4$ & 2.5:Mn$_3$CuN  & 2.54:Sr$_2$Cr$_3$As$_2$O$_2$\\
        \hline
        2.57:TbMn$_2$Si$_2$ & 2.59:Mn$_3$As$_2$ & 2.65:UPd$_2$Si$_2$  & 2.70:GdMg  & 2.77:Eu$_2$CuO$_4$  & 2.81:ErMn$_2$Si$_2$ & 2.82:ErMn$_2$Si$_2$ & 2.83:ErMn$_2$Ge$_2$  & 2.84:ErMn$_2$Ge$_2$ & 3.10:NpSe\\
        \hline
        3.11:NpTe  & 3.12:USb & 3.6:DyCu   & 3.7:NpBi  & 3.8:NdZn  & 3.9:NpS  &        &        &        & \\
        \hline
        \hline
\end{tabular}}
\caption{\textbf{The 236 magnetic topological materials predicted to be topological for at least 3 values of $U$ at the intrinsic filling.} ``ID'' denotes the ID in MAGNDATA  \cite{MT-SM-magndata}, and ``formula'' denotes chemical formula. We highlight magnetic materials with Kagom$\acute{\mathrm{e}}$ sublattice by printing the chemical formula in red.}\label{table-1}
\end{table*}

\begin{figure*}[!hbtp]
	\includegraphics[width=1\textwidth]{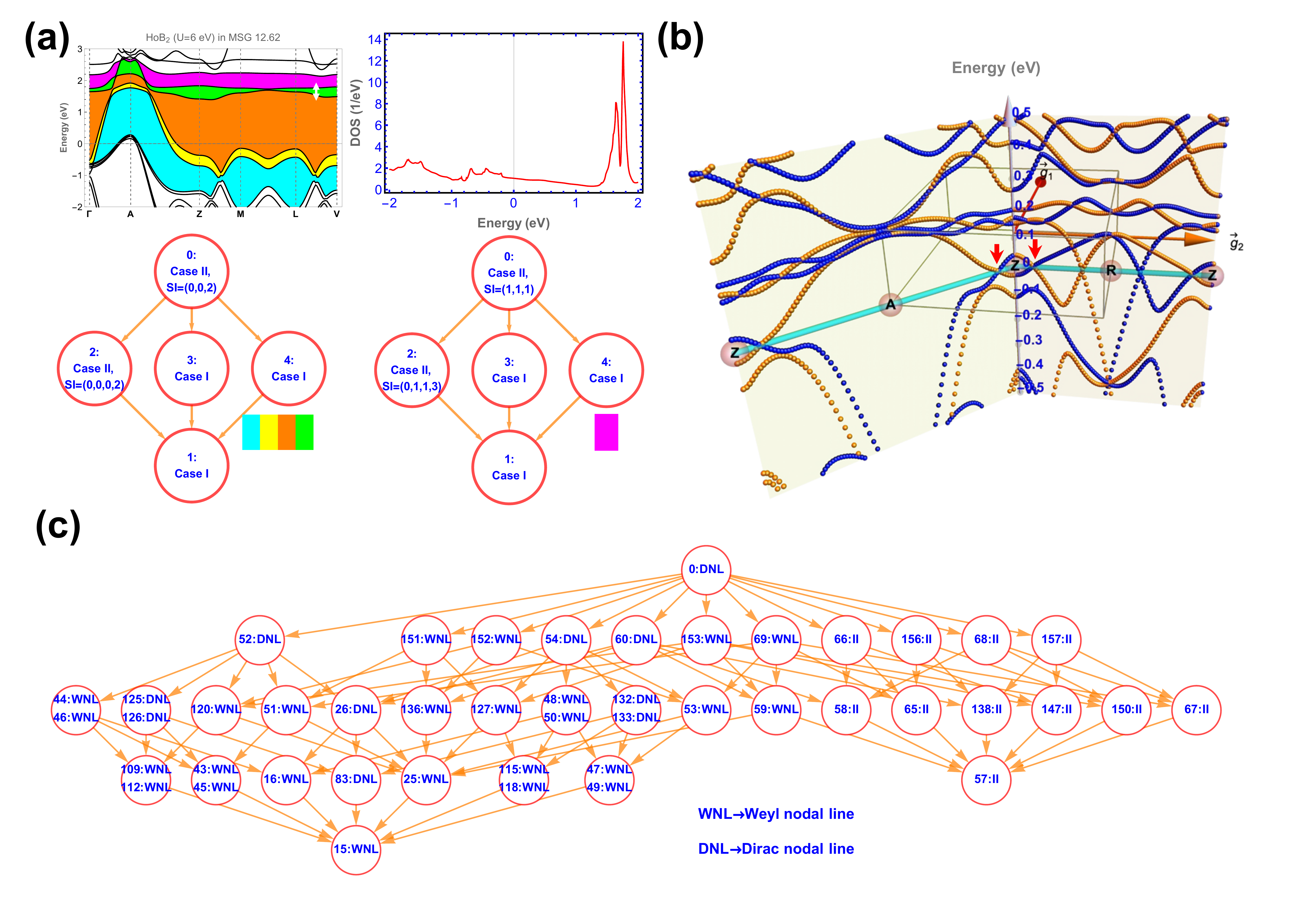}\\
	\caption{\textbf{TSEs in HoB$_2$ and CeMn$_2$Si$_2$.} In the upper panel of (a), we show the band structure   where all the bands are nondegenerate and the plot of DOS for HoB$_2$. The continuous gaps corresponding to the intrinsic filling $\nu_0$ and the fillings $\nu_0-1,\nu_0+1,\nu_0+2,\nu_0+3$ are indicated by the regions in yellow, cyan, orange, green and magenta, respectively.  The lower panel of (a) shows the tree-like graphs for the TSE corresponding to the gaps indicated by the  colors shown in the insets.  The symmetry-breaking patterns:  0, 1, 2, 3 and 4, correspond to the t-subgroups being MSGs 12.62, 1.1, 2.4, 5.15 and 8.34,  respectively.  Note that the peaks  in DOS correspond to the two bands (the band indices are $\nu_0+2$ and $\nu_0+3$) indicated by the white double-arrow. Such quasi-flat bands are found to be topological. In (b), we show the band structures for CeMn$_2$Si$_2$ along  two HSLs, $(0,w,1/2)$ and $(w,w,1/2)$, where the orange and blue colors encode two different co-irreps of the corresponding HSLs. Various BCs composed of these co-irreps can be found and two BCs indicated by the red arrows are deduced by the symmetry-data at A and Z, and  the symmetry-data Z and R, respectively (see Eq. \ref{CR2}). Note that in each of these HSLs, there exists another BC dictated by an MSG operation.
(c) shows the tree-like graph for the evolution of the  original Dirac  nodal lines  in CeMn$_2$Si$_2$. In each circle, the integer denotes  the symmetry-breaking pattern, II stands for case II and WNL/DNL stands for
Weyl nodal line/Dirac nodal line in case III.
}\label{figure-3}
\end{figure*}

We also manually select 9 magnetic topological materials whose electronic band plots are depicted in Fig. \ref{figure-3},  owning either  a noticeable band gap or  enforced BC almost at the Fermi level. The materials in Fig. \ref{figure-3} are topological for at least four values of $U$ except Sr$_2$TbIrO$_6$, for which only two jobs ($U = $0, 4 eV) reach convergence.  Roughly speaking, topological classification being case II/III definitely means the material is topological while case I means that the material is possibly topologically trivial \cite{MT-SM-Tang-NP,Tang-SA}. See Sec. S2 of SM \cite{MT-seeSM} for details on case I/II/III.   Note that different values of $U$  might correspond to different  topological classifications (case II/III) in UAsS and Ba$_3$CoIr$_2$O$_9$.  For others, different values of $U$  give the same classification of either case II or case III.

By symmetry-breaking,  the nontrivial band topology in these materials can be maintained, transformed or trivialized. For example, the topological classifications for Sr$_2$TbIrO$_6$ with $U = 0$ and 4 eV  are identical, with respect to all t-subgroups (including the parent MSG): for the parent group, SI $=(1)\in\mathbb{Z}_2$ and for t-subgroup MSG 2.4, SI $=$ $(0,0,0,2)\in\mathbb{Z}_2\times\mathbb{Z}_2\times\mathbb{Z}_2\times\mathbb{Z}_4$ \cite{MT-SM-M-SI} while for other t-subgroups, the classifications are case I. Besides, 21 of the 35 t-subgroups  for UAsS ($U$ $=$ 0, 2, 4, 6 eV) give case II/case III prediction, indicating the robustness of nontrivial band topology against symmetry-breaking.  For CaMnSi ($U$ $=$ 0, 1, 2, 3 eV), 5 of the 35 t-subgroups give case II while the rest give case I. For CsMnF$_4$ with  $U = 0, 1, 2,  3$ eV,  the topological classifications share the same result with respect to all t-subgroups.  For this material, 8 of the 16 t-subgroups give case III prediction and the rest give case I, and the enforced BCs include two pairs of opposite-charged iso-energetic Weyl points and one Weyl nodal line for the parent MSG. The Weyl points of each pair are related by inversion symmetry. By symmetry-breaking, the original Weyl points/nodal line can be gapped: For example, with respect to the t-subgroup MSG 6.18, the original Weyl points are all gapped while the original Weyl nodal line still exists); With respect to the t-subgroup MSG 18.19, the original inversion related Weyl points are allowed to own different energies since the inversion symmetry is broken in this t-subgroup.  For the 9 materials in Fig. \ref{figure-3}, we provide details in Sec. S3 of SM \cite{MT-seeSM} to demonstrate the diagnosis of band topology combining MAGNDATASymData and CalTopoEvol.

\begin{figure*}[!bhtp]
	\includegraphics[width=1\textwidth]{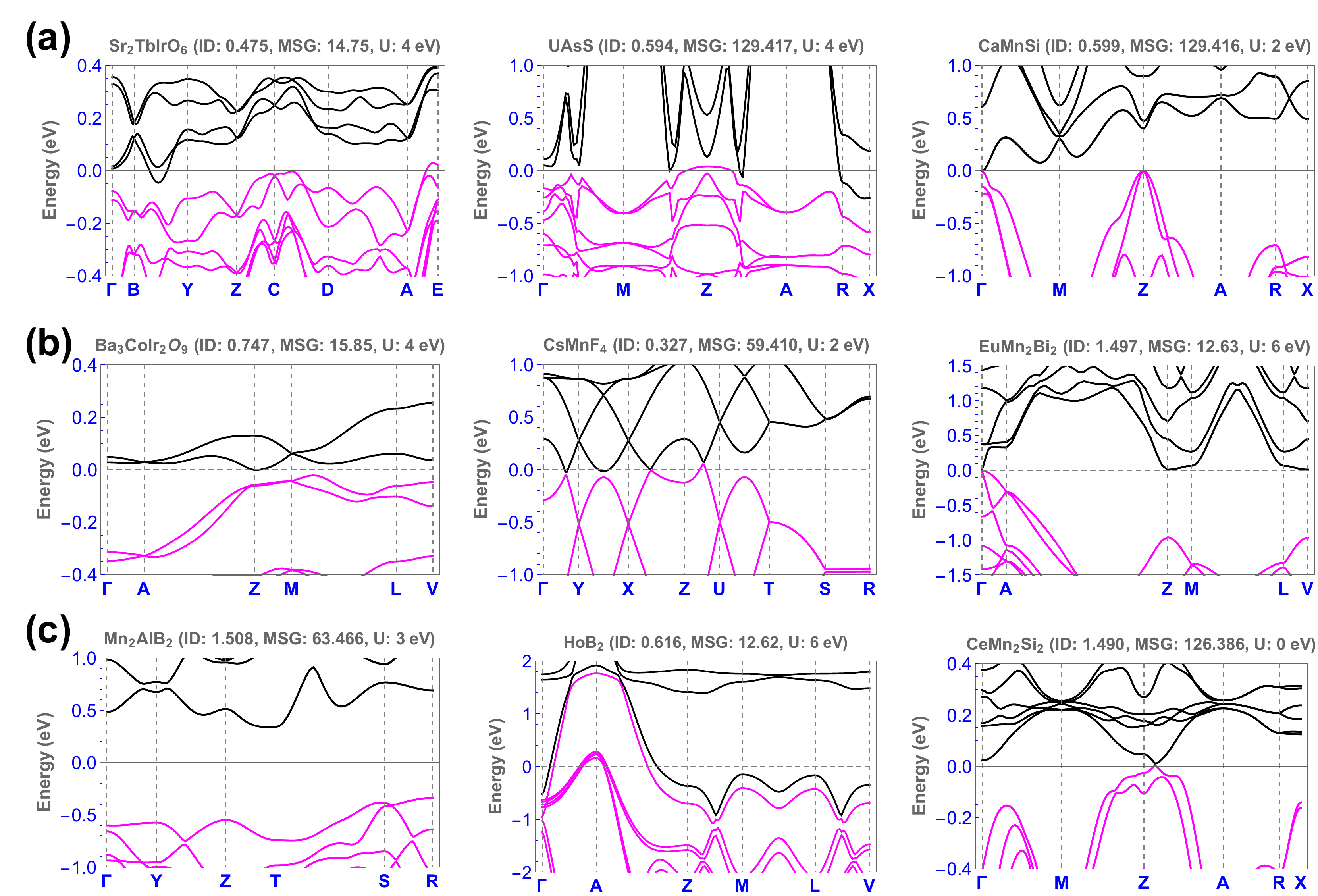}\\
	\caption{\textbf{Energy band plots of nine selected magnetic topological materials identified  in this work.} In all the band plots, the Fermi level is set to be 0 eV. The chemical formula, the ID in MAGNDATA \cite{MT-SM-magndata}, the MSG and the Hubbard $U$ are all provided in the head of each band plot. We consider the intrinsic filling $\nu_0$ and the bands with indices lower than or equal to $\nu_0$ are printed in  magenta. We find that CsMnF$_4$  and CeMn$_2$Si$_2$ are classified in case III owning enforced BCs. The rest are all in case II owning nonvanishing values in the corresponding SI groups.}\label{figure-2}
\end{figure*}

\section{Examples}\label{Mat-example}

In the following, we take HoB$_2$ and CeMn$_2$Si$_2$ as examples to show the detailed TSEs along the continuous symmetry-breaking paths.
\subsection{Materials example: HoB$_2$}
HoB$_2$ is classified to belong to case II. The ID in MAGNDATA \cite{MT-SM-magndata} is 0.616. HoB$_2$ in the paramagnetic state is crystallized in a hexagonal lattice with a very simple crystal structure containing only one chemical formula per primitive unit cell,  while a ferromagnetic state breaks the three-fold rotation symmetry, resulting in  MSG 12.62, ever reported to display a gigantic magnetocaloric effect \cite{MT-HoB2,MT-HoB2-2}. The resulting magnetic point group consists of $E, I, \Theta C_{2z}$ and $\Theta \sigma_{z}$ where $E$ and $I$ are unitary operations (identity and spatial inversion, respectively). The other two are antiunitary operations ($\Theta$ denotes the time-reversal operation, $C_{2z}$ is the 2-fold rotation around $z$-axis and $\sigma_z$ is a mirror operation).   To list all t-subgroups, one only need to enumerate all subgroups of the magnetic point group, as follows (in the form of ``symmetry-breaking pattern$\rightarrow$ preserved operations''):
\begin{equation}\notag
\begin{aligned}
  &0\rightarrow \{E,I,\Theta C_{2z},\Theta \sigma_{2z}\},\\
  &1\rightarrow \{E\},\\
  &2\rightarrow \{E,I\},\\
  &3\rightarrow \{E,\Theta C_{2z}\},\\
  &4\rightarrow \{E,\Theta \sigma_{2z}\},
\end{aligned}
\end{equation}
Besides, for MSG 12.62, all energy bands should be non-degenerate by symmetry requirements, and also, BCs are not allowed to exist stably at the special $k$ points. MSG 12.62 owns a nontrivial SI group \cite{MT-SM-M-SI}: $\mathbb{Z}_2\times\mathbb{Z}_2\times\mathbb{Z}_4$. Indeed, for all t-subgroups, HoB$_2$ can only be case I or case II as shown for different fillings in the lower panel of Fig. \ref{figure-3}(a) ($\nu=\nu_0,\nu_0-1,\nu_0+1,\nu_0+2$ and $\nu_0+3$) which characterize the TSEs by the tree-like graphs.  The fillings correspond to the gaps shown in the upper panel of Fig. \ref{figure-3}(a), indicated by different colors. Note that there are considerable  gaps for the filling $\nu_0+1$ (the region in orange in the upper panel of Fig. \ref{figure-3}(a)), the nontrivial topology of which is expected to bring about noticeable protected boundary states.  Such gaps share the same TSEs as those for another three fillings as shown in the lower panel of Fig. \ref{figure-3}(a), where the colors shown in the inset correspond to the gaps in the respective colors of the band plot. Interestingly, the quasi-flat bands (whose band indices are $\nu_0+2$ and $\nu_0+3$) are found to topological. The coexistence of magnetic order, topological quasi-flat bands  and multiple topological gaps around the Fermi level makes this material a fascinating platform to study possible exotic quantum excitations.

\subsection{Materials example: CeMn$_2$Si$_2$}
We then show another example for CeMn$_2$Si$_2$ whose ID in MAGNDATA is 1.490 \cite{MT-SM-magndata} and MSG is 126.386. The Ce ion is found to own vanishing magnetic moment while the magnitude of the magnetic moments in Mn ions are found to be 1.9 $\mu$B  experimentally \cite{MT-CeMn2Si2}.  Here we chose the results for the calculations in which all the values of $U$ on the Ce and Mn ions  are  $0$ eV, in which the calculated magnetic moments reasonably reproduce those by experiments.
It is noted that the other results for the rest values of $U$ share similar topological properties discussed here. Consider the intrinsic filling. With respect to the parent MSG, CeMn$_2$Si$_2$ is predicted to be in case III and furthermore, the HSP symmetry-data guarantee that the enforced BCs appear in HSLs,  $(0,w,\frac{1}{2})$ and $(w,w,\frac{1}{2})$, as shown in Fig. \ref{figure-3}(b), where these BCs are also verified from the computed representations of a dense $k$-point mesh  in the HSLs. The BCs in these HSLs actually lie in a Dirac nodal line within an HSPL ($(u,v,\frac{1}{2})$) by CRs. Then consider the effect of the symmetry-breaking. These Dirac nodal lines can  be gapped for some symmetry-breaking patterns,  resulting in a possibly-trivial insulator (in case I) or a topological phase in case II. For other symmetry-breaking patterns, the resulting topological phases are all in case III, and furthermore, we find that the corresponding enforced BC lies in either Dirac nodal line or Weyl nodal line. We collect the TSEs for this material in Fig. \ref{figure-3}(c) along all symmetry-breaking paths containing the symmetry-breaking patterns with topological classifications being case II/III.
\section{Contraction of tree-like graph}
\begin{figure*}[!t]
	\includegraphics[width=1\textwidth]{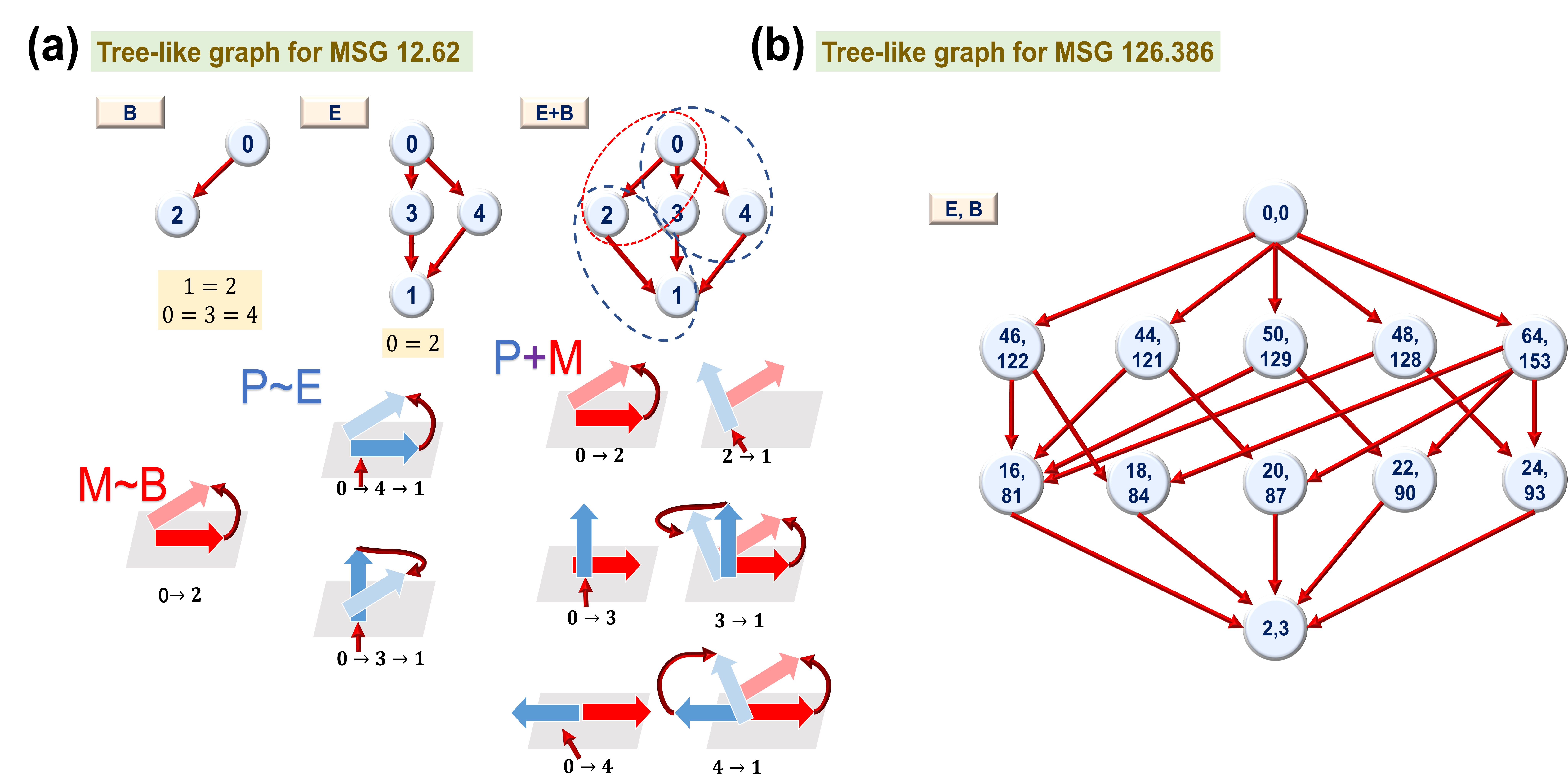}\\
	\caption{\textbf{Contraction of tree-like graph in MSGs 12.62 and 126.386.} (a) Consider MSG 12.62. EBSToSubgrps can quickly output the contracted tree-like graph by applying $\vec{B}$ or $\vec{E}$ or simultaneously applying  $\vec{E}$ and $\vec{B}$ as shown here. Concretely, applying $\vec{B}$ would result in t-subgroups by 0 and 2 ($\vec{B}=(B_x,B_y,0)$ and $(B_x,B_y,B_z)$, respectively). Applying $\vec{E}$ would result in t-subgroups by 0, 1, 3 and 4 ($\vec{E}=(0,0,0), (E_x,E_y,E_z)$, $(0,0,E_z)$ and $(E_x,E_y,0)$, respectively). Simultaneously applying $\vec{E}$ and $\vec{B}$ result in all t-subgroups, $0, 1, 2, 3$ and $4$ ($\{\vec{E},\vec{B}\}=\{(0,0,0),(B_x,B_y,0)\}, \{(E_x,E_y,E_z),(B_x,B_y,B_z)\}$, $\{(0,0,0),(B_x,B_y,B_z)\}, \{(0,0,E_z),(B_x,B_y,0)\}$ and $\{(E_x,E_y,0),(B_x,B_y,0)\}$, respectively). The contractions shown by blue and red dashed ellipses in the tree-like graph labeled by ``E+B'' results in different resulting tree-like graphs labeled by ``B'' and ``E'', respectively.  In the lower panel, we show that the continuous switching of macroscopic electric polarization  $\vec{P}$ or/and magnetization  $\vec{M}$ for each tree-like graph, demonstrating their emergence accompanying the symmetry-breaking (and TSE). (b)   MSG 126.386 owns in total 158 t-subgroups, and 1581 symmetry-breaking paths from symmetry-breaking pattern 0 (the parent MSG) to symmetry-breaking pattern 1 (P1), while applying $\vec{E}$ or $\vec{B}$, only partial subgroups  can be realized and are shown here. The corresponding tree-like graphs applying $\vec{E}$ and $\vec{B}$ share the same form but contain different t-subgroups as indicated in the vertexes (say, $a,b$ means that $a$ denotes the t-subgroup for $\vec{E}$ and $b$ represents the t-subgroup for $\vec{B}$, respectively).  Detailed symmetry operations that are presented by various fields can be found by EBSToSubgrps.}\label{figure-4}
\end{figure*}

We then discuss an interesting finding that the tree-like graphs could be contracted in  different patterns when we specify different concrete realizations of symmetry-breaking. As has been schematically shown in Fig. \ref{figure-1}(b), the general tree-like graph shown in Fig. \ref{figure-1}(a) could transform into another different patterns once for the symmetry-breaking patterns being 0, 2, 3 and 4 (or for those being 4, 5 and 6), the concrete  realizations of symmetry-breaking take the same form (e.g. electric fields are all long $z$ direction). Take MSG 12.62 and its t-subgroups as an example. Once we only apply magnetic field, $\vec{B}$ (or a variation of magnetization $\vec{M}\sim \vec{B}$), only the t-subgroups with symmetry-breaking patterns 0 and 2 can be realized, as shown in Fig. \ref{figure-4}(a). Once we only apply electric field, $\vec{E}$ (or an electric polarization $\vec{P}\sim\vec{E}$), only the t-subgroups with symmetry-breaking patterns 0, 3, 4 and 1 can be realized, as shown in Fig. \ref{figure-4}(a). Only when we consider the aforementioned two types of fields/orderings simultaneously, all t-subgroups can be realized.  As shown in the lower inset of Fig. \ref{figure-4}(a), different orderings can lead to different switching patterns along all symmetry-breaking paths. For HoB$_2$ as described above, we can infer that once the nontrivial band topology is preserved in a process of phase transition,  a macroscopic electric polarization is not sufficient since it could break inversion symmetry that protects the topology. In Fig. \ref{figure-4}(b), we show another example using MSG 126.386, that applying $\vec{E}$ and $\vec{B}$ would give rise to the same tree-like graph pattern but the detailed t-subgroups in the vertexes of the tree-like graph are totally different. From Fig. \ref{figure-3}(c), one can conclude that applying $\vec{E}$ would gap the original Dirac lines or transform them to Weyl nodal lines. However, applying $\vec{B}$ would definitely gap these Dirac nodal lines.
\section{Conclusion and perspective}
To conclude, the three subroutines obtained in this work are expected guide experimentalists to choose suitable material and specify concrete symmetry-breaking, which might pave the avenue to the realization of highly-sensitive magnetism-controlled band topology.  The high-throughput calculation results can also be used as the training dataset to develop a simple-to-use heuristic chemical rule of diagnosing band topology such as the topogivity \cite{MT-ML} by machine-learning. Note that the predicted TSE is irrelevant with the origin of concrete symmetry-breaking, while a concrete symmetry-breaking leading to a contraction of the tree-like graph relates TSE with the symmetry-breaking origin. Similar scheme could be established for superconductors
\cite{MT-SM-Ono-PRR,MT-SM-Tang-TNSC}.
Note that the developed subroutines CalTopoEvol and EBSToSubgrps can be applied to both nonmagnetic and magnetic materials. It is also worth pointing out that detailed band topology designating where the protected boundary states can be found can be obtained by checking the results as listed in Refs. \cite{MTQC, Peng}.

The correlation of magnetism and nontrivial band topology  with other ordering, such as ferroelectric order \cite{MT-ferro-physics}, and novel crystal structure (e.g. Kagom$\acute{\mathrm{e}}$ lattice \cite{Kagome-review}, and see Table \ref{table-1} and Sec. 8 of SM \cite{MT-seeSM} for selected magnetic materials with Kagom$\acute{\mathrm{e}}$ sublattice), merits future studies by choosing suitable candidates from our predicted magnetic topological materials.   Besides, a more generic symmetry-breaking could include the translation symmetry breaking  enlarging the primitive unit cell and lead to Brillouin zone folding (e.g. a charge-density wave order \cite{MT-FeGe, SM-FeGe-2}), which is not  exhaustible, but can still apply the t-subgroups  after firstly identifying the maximal subgroup(s),  of which the details  are described in Sec. S5 of SM \cite{MT-seeSM}.

\section{Acknowledgments}
We are very grateful for earlier collaborations on related topics with Ashvin Vishwanath, Hoi Chun Po, Haruki Watanabe and Seishiro Ono.  F.T. appreciates insightful discussions with Wei Chen, Qun-Li Lei, Kai Li and Yang-Yang Lv. We also thank very helpful suggestions from Ge Yao on high-performance computing. \textbf{Funding: }F.T. was supported by National  Natural  Science Foundation of China (NSFC) under Grant No. 12104215, No. 12322404 and Young Elite Scientists Sponsorship Program by China Association for Science and Technology. F.T. and X.W. were supported by  NSFC Grants No. 12188101,  No. 11834006, the National Key
R\&D Program of China (Grant No.2022YFA1403601) and
Innovation Program for Quantum Science and Technology,
No. 2021ZD0301902. X.W. also acknowledges the support from the Tencent Foundation through the XPLORER PRIZE and Natural Science Foundation of Jiangsu Province (No. BK20233001, BK20243011).
\textbf{Author contributions:} F.T. conceived and designed the project, and performed all  calculations. All authors contributed to the writing and editing of the manuscript.
\textbf{Competing interests:} None declared.
\textbf{Data and materials availability:} The subroutines, CalTopoEvol, MAGNDATASymData, EBSToSubgrps can be downlocaed at, \url{https://github.com/FengTang1990/CalTopoEvol}, \url{https://github.com/FengTang1990/MAGNDATASymData}, and \url{https://github.com/FengTang1990/ExternalFields}, respectively.
\bibliography{Refs}
\end{document}